\documentclass[preprint,prb,showpacs,preprintnumbers,amsmath,amssymb]{revtex4}
\usepackage{epsfig,amsmath,graphicx,amssymb}
\usepackage{float}
\begin{document}
\title{Polarization-resolved cartography of light emission of a VCSEL with high space and frequency resolution}

\author{Tao Wang$^{\rm 1,2,a}$ and Gian-Luca Lippi$^{\rm 1,2,b}$}
\affiliation{
$^{\rm 1}$ \mbox{Institut Non Lin\'eaire de Nice, Universit\'e de Nice Sophia Antipolis}\\
$^{\rm 2}$ \mbox{CNRS, UMR 7335, 1361 Route des Lucioles, F-06560 Valbonne, France}\\
$^{\rm a}$ email: taowang166@gmail.com\\
$^{\rm b}$ email: Gian-Luca.Lippi@inln.cnrs.fr}

\date{\today}

\begin{abstract}
We couple a double-channel imaging technique, allowing for the simultaneous acquisition of high-quality and high-resolution intensity and peak emission wavelength profiles [T. Wang and G.L. Lippi, Rev. Sci. Instr. {\bf 86}, 063111 (2015)], to the polarization-resolved analysis of the optical emission of a multimode VCSEL.  Detailed information on the local wavelength shifts between the two polarized components and on the wavelength gradients can be easily gathered.  A polarization- and position-resolved energy balance can be constructed for each wavelength component, allowing in a simple way for a direct analysis of the collected light.  Applications to samples, other than VCSELs, are suggested.
\end{abstract}

\maketitle

Current technology relies heavily on the high manufacturing quality of light-emitting devices, thus on beam quality (spatial and emission wavelength homogeneity) and on polarization control.  Among the various light sources, Vertical-Cavity Surface-Emitting Lasers (VCSELs) stand out because of their single longitudinal mode operation, small beam divergence, (nearly) circular spot profile, very low threshold current and ease of integrability.  The excellent symmetry properties currently obtained in VCSEL cavity structure, contrary to the strong asymmetry between TE and TM modes intrinsic to edge emitters, relinquish all control on the polarization state of the emitted electromagnetic (e.m.) field -- manufacturing details and imperfections become therefore of paramount importance and generate complex polarization dynamics~\cite{Doorn1996,Mukaihara1993,Thompson1997,Debernardi2002,Sondermann2003,Sondermann2004} where the emission on one linearly polarized component can switch to the orthogonal direction~\cite{Hasnain1991,Choquette1994}.  The need for polarization characterization and control~\cite{Panajotov2001,Michalzik2012} still drives research and development in VCSEL construction. 

In multi-transverse mode VCSELs small frequency splittings, originating from intrinsic birefringence, have been identified as a source of polarization switching~\cite{Balle1999} and the minimum difference in refractive index between the two components, necessary to stabilize operation in one polarization, has been estimated~\cite{Valle1996} to be of the order of $5 \times 10^{-5}$.  Experimental measurements have confirmed this finding~\cite{Doorn1994,Exter1997,Yu2013,Zhang2014} and have paved the way for designing polarization-stable devices~\cite{Choquette1995,Hendriks1997,Park2000}.  In the past, numerous techniques~\cite{Chavezpirson1993,Berger1996,Doorn1996,Hao2011} have been devised, including interferometric (Fabry-Perot) measurements~\cite{Hendriks1997}, reflectance difference spectroscopy~\cite{Yu2012,Yu2013} and near-field imaging~\cite{Hoersch1996,Rhodes1998,Yu2007,Fratta2001,Debernardi2009}, but most of them are tailored to a specific kind of sample or to particular experimental conditions.  Thus, there is still a need for a more general and flexible measurement technique.

\begin{figure}[H]
\centering
\includegraphics[width=0.9\linewidth,clip=true]{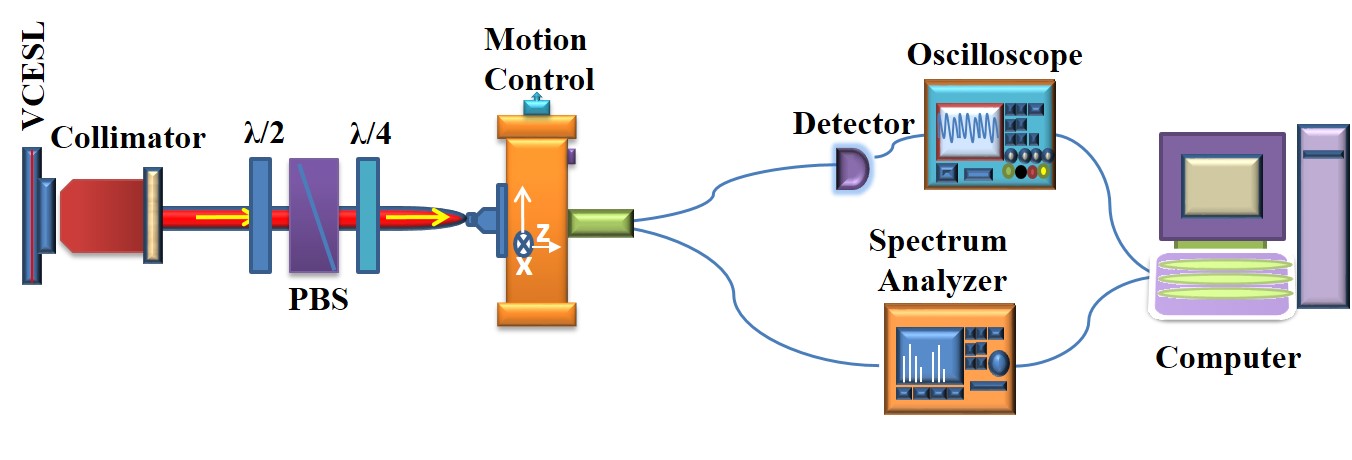}
\caption{
Experimental setup.  The collimator (Thorlabs, C230TM-B) is adjusted to obtain the near-field image in the scanning plane of the profiler~\cite{Tao2015} and is preferred here to a microscope objective given the minimal distance required to insert the various optical elements (wave plate and PBS).  The $\lambda/2$ plate is adjusted to select either polarization component (and the ensemble $\lambda/2$-plate--PBS is removed for unpolarized images).  The other key elements -- photodetector (high-gain UDT-455), the spectrum analyzer (Agilent 86142B) and the digitizing oscilloscope (Agilent DSO-X 3024A) -- are those already used~\cite{Tao2015}.  Measurements are taken with the spatial resolution chosen for the unpolarized scans~\cite{Tao2015}.
}
\label{Experimentalsetup}
\end{figure}

In this letter, we take up this challenge and present a scheme allowing for the simultaneous, high-resolution, polarization-resolved cartographic reconstruction of the intensity and frequency distribution of emitted e.m. field intensity.  The direct imaging of the emission frequency, associated to the intensity pattern and to the polarization information, allows for an immediate pictorial description of the emission and helps in identifying and quantifying the device's multi-mode features. 

As a test device, we use a multimode, commercial semiconductor laser emitting at $\lambda = (980\pm3) nm$ (Thorlabs VCSEL--980) with maximum output power $P_{max}(i \approx 10 mA) = 1.85mW$.  The measured threshold current is $i_{th} \approx 1.3 mA$ and the estimated aperture diameter is $d \approx 6 \mu$m (deduced from the manufacturer's datasheet). The laser is mounted on a temperature-stabilized holder (Thorlabs TCLDM9) with stability better than $0.1K$ and is supplied by a commercial source (Thorlabs LDC200VCSEL) with resolution $1 \mu A$ and accuracy $\pm 20 \mu A$.  The polarization analysis is performed by inserting a $\lambda/2$-plate, followed by a Polarizing Beam Splitter (PBS), in the beam path of a scannning device~\cite{Tao2015} which samples the image of the near-field produced by the collimator (Fig.~\ref{Experimentalsetup}).  
The total average laser power output as a function of injection current and the typical optical spectrum at $i = 5 mA$ are displayed in Fig.~\ref{IVcurve,spectrum}, which shows: (a) the lasing threshold at $i_{th} \approx 1.3 mA$, and (b) the emission wavelength centered between $978$ and $979 nm$.

\begin{figure}[H]
\centering
\includegraphics[width=1\linewidth,clip=true]{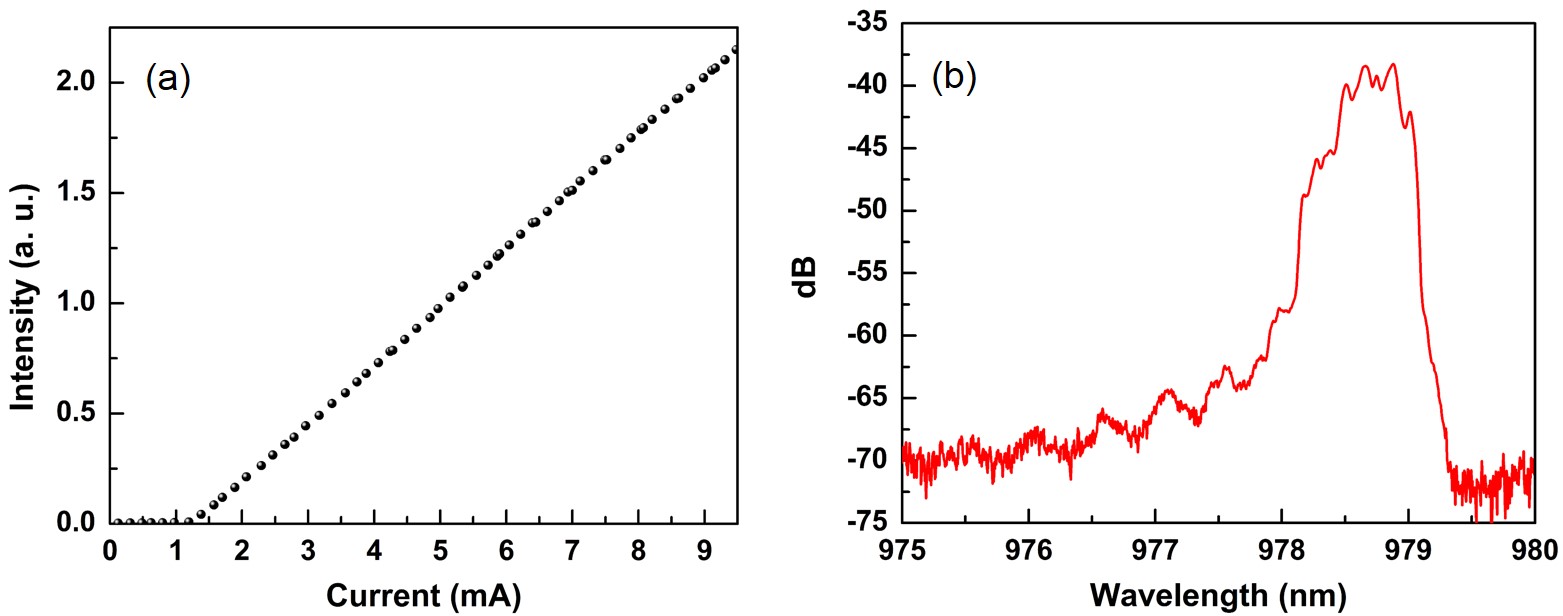}
\caption{Unpolarized input-output laser characteristics (a) and optical spectrum at $i=5 mA$ (b).}
\label{IVcurve,spectrum}
\end{figure}

The selected, inexpensive multimode VCSEL is well suited for a detailed cartographic demonstration due to its quite complex modal superposition leading to a rich spectral distribution in the total and polarization-resolved output.
Fig.~\ref{5mAbeam,5mAwave2}(a) shows the typical near-field intensity distribution emitted by the VCSEL at $i=5 mA$.  A three-fold symmetry is apparent, with a strong localised emission peak in one of the lobes (cf. in particular the 3D inset).
While the overall symmetry is likely to originate from the superposition of angularly symmetric (Ince-Gauss) modes~\cite{Bandres2004}, the ``hot spot" is probably due to current crowding.
Fig.~\ref{5mAbeam,5mAwave2}(b) shows the associated, color-coded, spatially resolved peak emission wavelength~\cite{Tao2015}. The spectral information complements the one contained in the intensity profile.  Comparison between the two panels (Fig.~\ref{5mAbeam,5mAwave2}, plotted on the same geometrical scale) shows that the two upper intensity lobes (panel (a)) share an emission wavelength  ($\lambda - \lambda_{ref} \approx -0.3 nm$), while the bottom one has its peak shifted further towards the blue  ($\lambda - \lambda_{ref} \approx -0.7 nm$).

\begin{figure}[H]
\centering
\includegraphics[width=1\linewidth,clip=true]{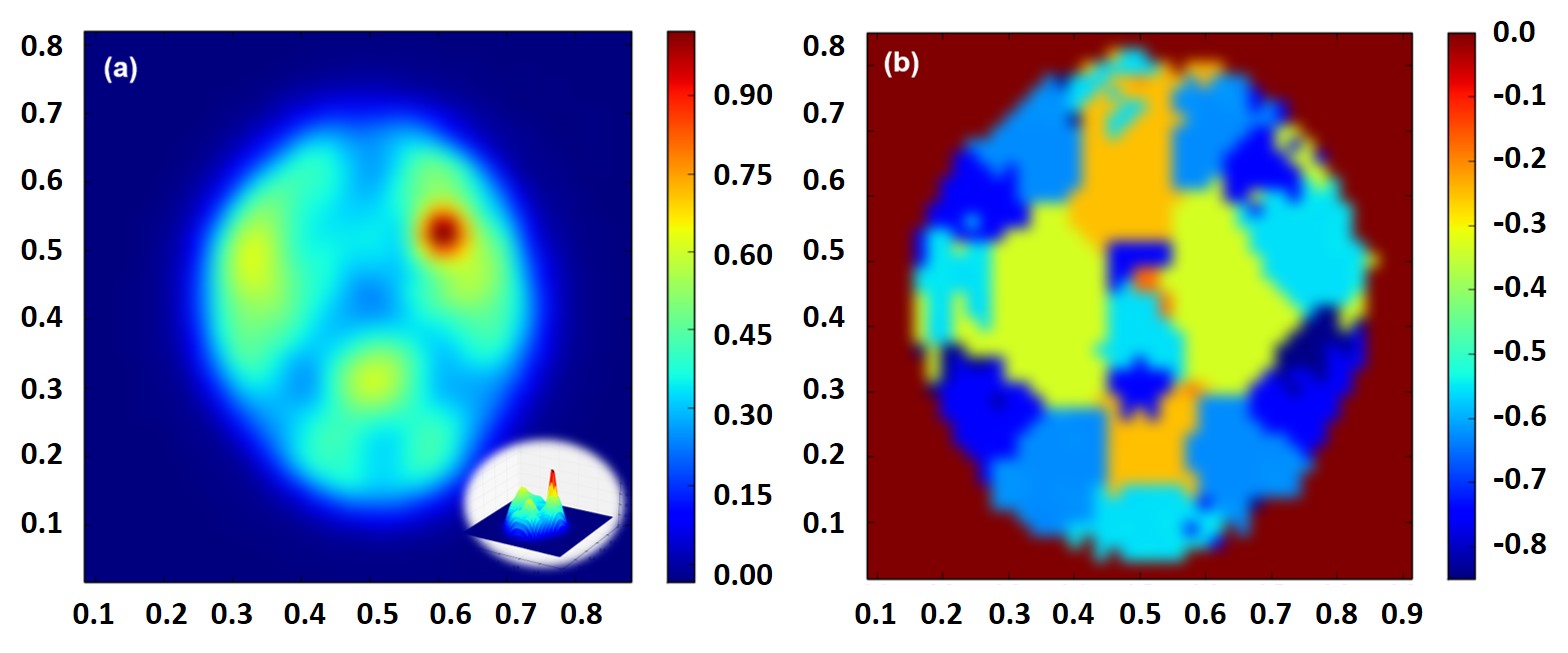}
\caption{Total mean output intensity and wavelength distribution profiles at pumping current $i=5 mA$: (a) intensity distribution profile; (b) spatial wavelength distribution.  All spectral distributions are plotted -- here and in the following -- relative to a reference wavelength $\lambda_{ref} = 979.2 nm$.  The color scale in the spectral pictures corresponds to the actual wavelength shift (blue representing the shortest wavelength components present in each spectrum).  The digit before the decimal point in each wavelength scale corresponds to $10^{-9} m$ (i.e. $nm$).
}
\label{5mAbeam,5mAwave2}
\end{figure}  

\begin{figure}[H]
\centering
\includegraphics[width=0.75\linewidth,clip=true]{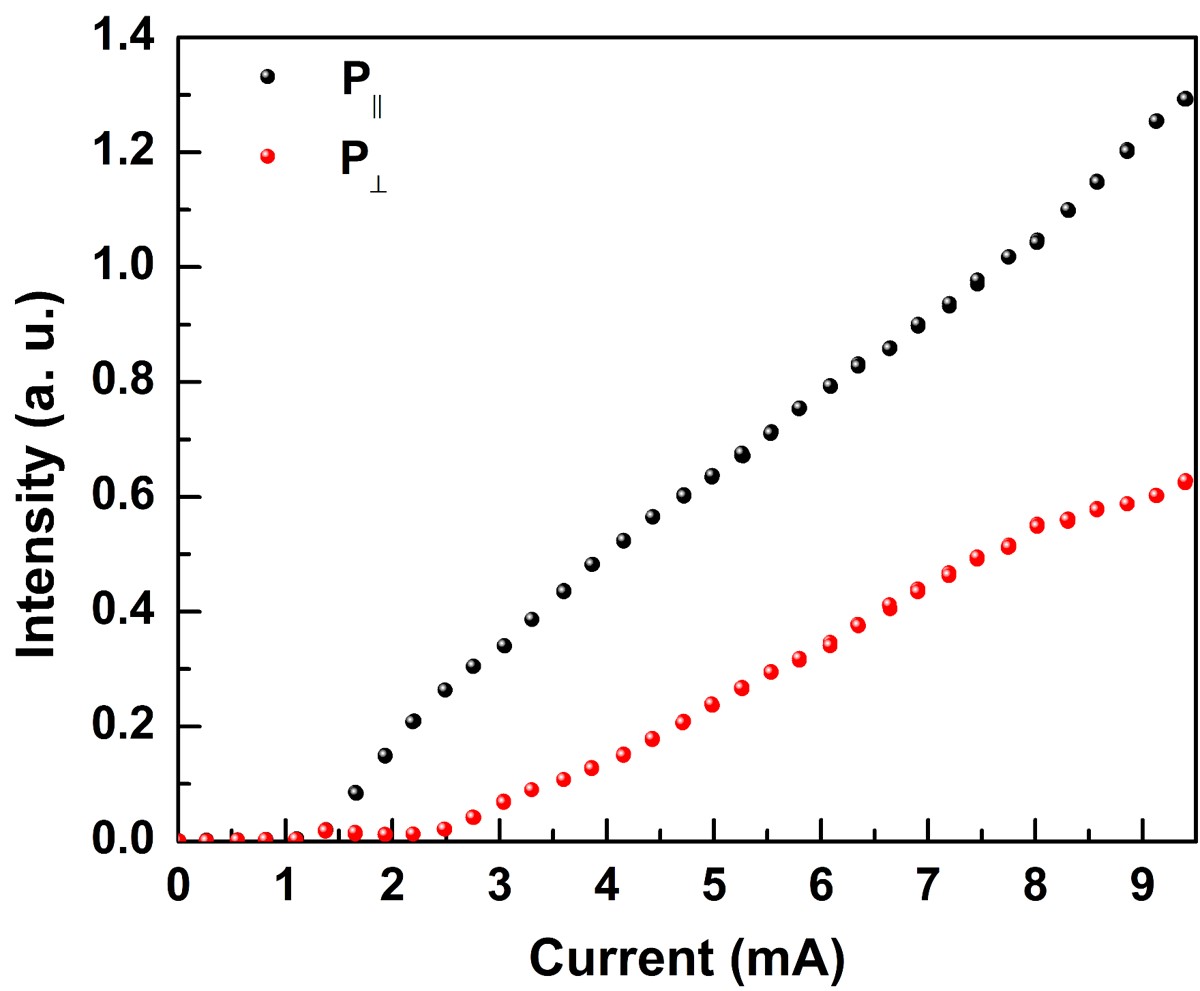}
\caption{Input-output lasing performance curves for the two $P_\parallel$ and $P_\perp$ polarization states.}
\label{PIV}
\end{figure}

The polarization-resolved input-output laser characteristics are displayed in Fig.~\ref{PIV} and show that, as expected, the weaker ($P_{\perp}$) component has a higher threshold.  However, at the operating current value, $i = 5 mA$, both components are active, with a power ratio $P_{\perp}:P_{\parallel} \approx 1:2.5$. 

\begin{figure}[H]
\centering
\includegraphics[width=1\linewidth,clip=true]{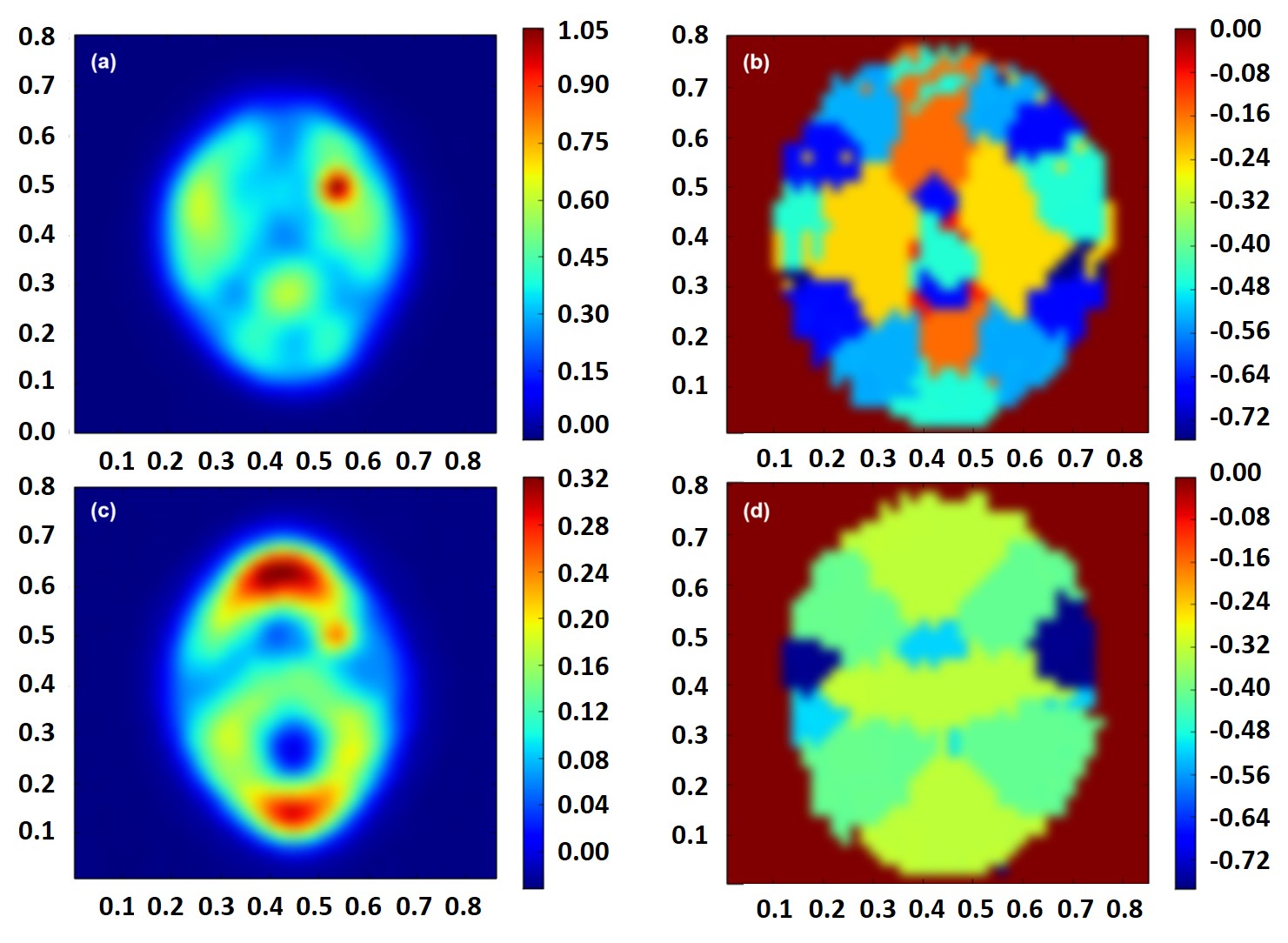}
\caption{Intensity and wavelength distribution for $P_\parallel$ (top) and $P_\perp$ (bottom) polarization states, separately. (a) and (c) are the intensity distribution profiles. (b) and (d) display the corresponding wavelength distributions.} 
\label{Duiying}
\end{figure}

The intensity patterns (Fig.~\ref{Duiying}(a,c)) show a polarization selective complementarity in the localization of respective maxima and minima, even though they do not lead to a complete smoothing of the total output (Fig.~\ref{5mAbeam,5mAwave2}(a)) due to the strongly unequal amount of power carried by the two polarizations (Fig.~\ref{PIV}). The spectral information, which thanks to the spectrometer's higher sensitivity extends beyond the range displayed by the intensity profile (as in Fig.~\ref{5mAbeam,5mAwave2}(b)), shows a much more homogeneous emission in the orthogonal, weak component ($P_{\perp}$, Fig. \ref{Duiying}(d)); its emission takes place at one main wavelengths ($\approx - 0.3 nm$ -- i.e., below $\lambda_{ref}$) and possesses a basic ``rectangular" symmetry.  The spectral information is much more complex in the principal polarization component ($P_{\parallel}$, Fig.~\ref{Duiying}(b)) where we see the appearence of stronger wavelength shifts in the beam wings with a more elaborate, rotationally symmetric structure.

\begin{figure}[H]
\centering
\includegraphics[width=0.8\linewidth,clip=true]{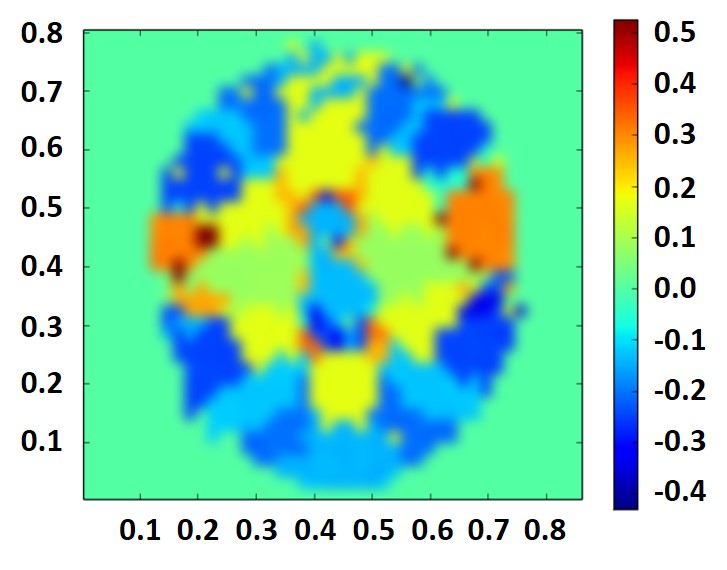}
\caption{Spatial distribution of wavelength difference ($\Delta \lambda_{\parallel,\perp} = \lambda_{P_{\parallel}} - \lambda_{P_\perp}$) in the overlap region for $P_\parallel$ and $P_\perp$ polarization states.
} 
\label{figuresb}
\end{figure}

The most interesting polarization-related information comes from the spatially resolved wavelength difference associated with the two polarized spectra (Fig.~\ref{figuresb}).  The background is now set at center scale ($0.00 nm$) while strong wavelength differences appear over the beam's cross-section ($\Delta \lambda_{\parallel,\perp} = 1 nm$).  Strong, isolated, small-sized regions where the emission wavelength strongly differs from its surrounding appear in the regions bordering the beam's ``edges" on the outer border ($\Delta \lambda_{\parallel,\perp} > 0$) as well as on the inner border ($\Delta \lambda_{\parallel,\perp} < 0$).  Their origin may be linked to material inhomogeneities, which would hardly influence the laser emission since this points are located in areas where the local intensity is very small (compare the coordinates of these points in Fig.~\ref{figuresb} and compare to Fig.~\ref{5mAbeam,5mAwave2}a); however, useful information about finer details (device structure, modal superposition, fabrication, etc.) may be carried by these spots.  Our interest here is simply in detecting and highlighting these regions, irrespective of their physical origin, to demonstrate the sensitivity and resolution of our technique.

The cartographic reconstruction well highlights the strong wavelength gradients which appear in the regions where the two polarized components switch their local intensity dominance.  Interestingly, areas of constant frequency difference between the two polarizations appear (e.g., the ``keyhole" shape in the center of the beam with wavelength difference $\Delta \lambda_{\parallel,\perp} \approx -0.15$), while the individual polarization spectra show a more complex pattern (Fig.~\ref{Duiying}b,d) and intensity gradients exist in the total (Fig.~\ref{5mAbeam,5mAwave2}a) and polarization-resolved (Fig.~\ref{Duiying}a,c) intensity distribution.    

Four ``outer" spots (located between 0.6 and 0.7 in the horizontal direction, and 0.4 and 0.6 in the vertical one in Fig.~\ref{figuresb}) stand out particularly well and show the presence of wavelength gradients up to $\frac{\Delta \lambda_{\parallel,\perp}}{\Delta s} \approx 0.5 nm / \mu m$ -- estimated from a wavelength difference $\Delta \lambda_{\parallel,\perp} \approx 0.1 nm$ over a spatial interval  $\Delta s \approx 0.2 \mu m$ (recostructed from the image on the basis an estimated diameter $d = 6 \mu m$ for the VCSEL).  Equating the relative wavelength variation to the relative refractive index change, we obtain an estimate (for GaAs, $n = 3.6$) for $\Delta n \approx 0.001$.

Notice that the wavelength resolution depends on the spectrometer.  Our current setup~\cite{Tao2015} has shown a peak wavelength sensitivity as low as $0.01 nm$, i.e., frequency resolution of 3 GHz (or 13 $\mu eV$ in energy units).  Better optical spectrum analyzers can further improve the performance.  This lower limit well matches the requirements for frequency-resolved measurements in VCSELs, whose birefringence-induced splitting has been found to range~\cite{Zhang2014} between $3 GHz$ and $22 GHz$.  It is therefore envisageable to use this technique to obtain spatially-resolved spectral information and reconstruct possible localised structural defects inducing birefringence in VCSELs with this technique.

The cartographic demonstration has been based on a multimode VCSEL, but the technique is suitable for many different samples.  For instance, photoluminescence measurements, routinely conducted in semiconductors~\cite{Bera2010} and thin films~\cite{Splendiani2010,Tongay2012,Gutierrez2013} , but also for monitoring the quality of large, centimeter-sized boules of semiconductor materials~\cite{Mitchell2014}, could benefit from the high spatial resolution, coupled to polarization and spectral information offered here.

\begin{figure}[H]
\centering
\includegraphics[width=0.7\linewidth,clip=true]{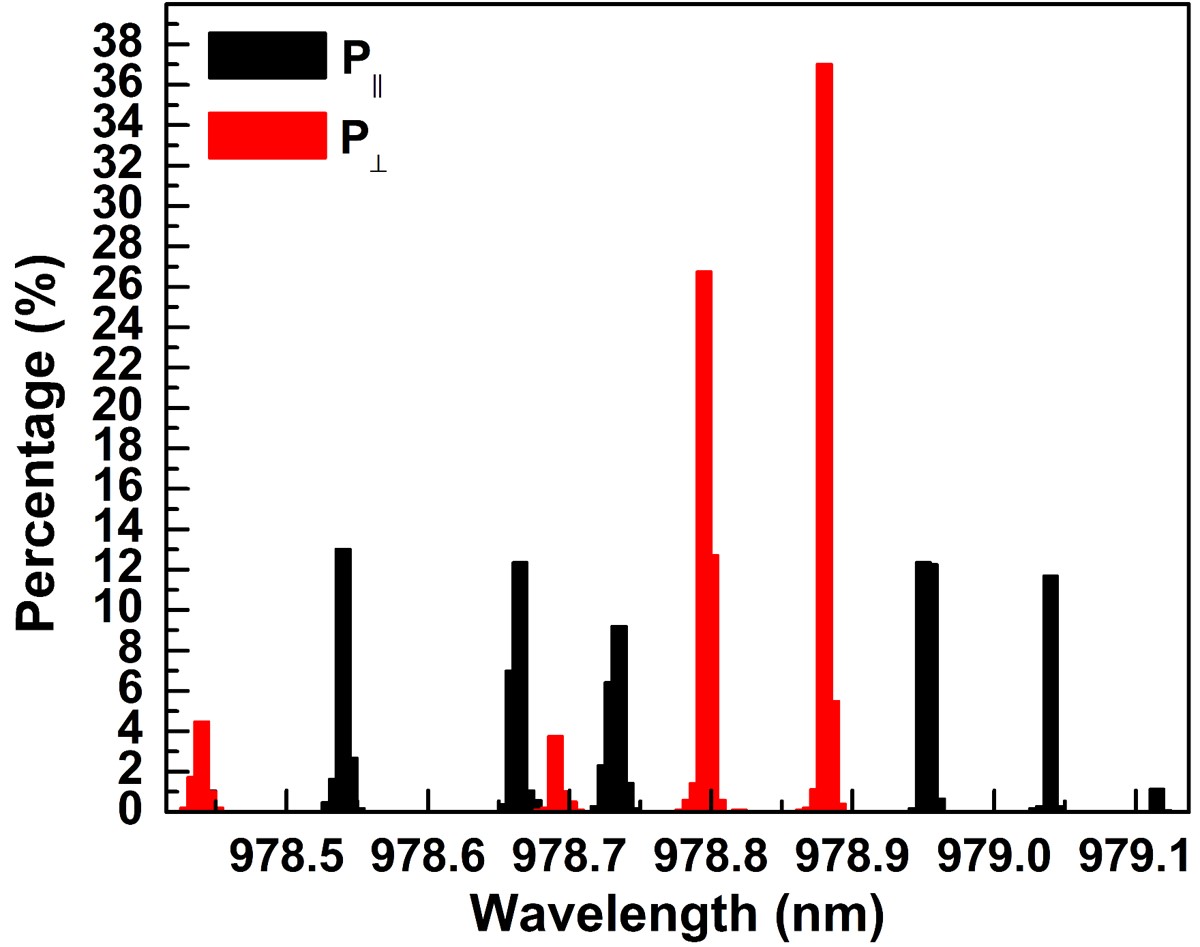}
\caption{Histogram of the number of ``pixels" (i.e. measured points in the sampled matrix) for each peak wavelength for $P_\parallel$ (black) and $P_\perp$ (red) polarization components.} 
\label{ratio}
\end{figure}

Before concluding, from Figs.~\ref{Duiying}(b,d) we  estimate the contribution of the different frequency components to the whole beam:   Fig.~\ref{ratio} distinctly shows the presence of different sets of discrete emission frequencies for the two polarized components.  This type of information, which provides statistical information on the relative fractions of the beam emitting at different wavelengths, can be straightforwardly completed by an energy balance (not shown):  adding up the intensity in each pixel, as a function of peak emission wavelength, one can end up with information equivalent to a peak spectral reconstruction, with the added advantage of the knowledge of the position from which each contribution is derived.  In other words, from this plot one can quantify the amount of energy carried at each frequency, in a fashion similar to an optical spectrum (e.g., Fig.~\ref{IVcurve,spectrum}(b)) where in addition one can also correlate the information to the geometrical location of each contribution.

In conclusion, we have demonstrated a cartographic technique which provides polarization- and spatially-resolved intensity and wavelength distributions of the emitted light with a resolution which can reach $3 GHz$.  A demonstration has been given specifically using a multimode VCSEL, but the technique can be applied to virtually any light-emitting source.

We are grateful to T. Ackemann, S. Barland, B. Garbin, L. Gil, M. Giudici, F. Gustave, M. Marconi for discussions and loan of instrumentation.  G. Almuneau and K. Panajotov are very warmly thanked for their advice on the manuscript and for physical discussions on VCSELs. Technical support from J.-C. Bery (mechanics) and from J.-C. Bernard and A. Dusaucy (electronics) is gratefully acknowledged. T.W. acknowledges a Ph.D. Thesis contract from the Conseil R\'egional PACA and support from BBright.

\end{document}